\documentclass[twocolumn,twoside,preprintnumbers,amsmath,amssymb,pacs]{revtex4}
\usepackage{epsfig}
\usepackage{graphicx}
\usepackage{dcolumn}
\usepackage{bm}

\usepackage{fancyhdr}

\usepackage{pslatex}

\newcommand{\half}{\mbox{\small{$\frac{1}{2}$}}} 
\newcommand{\Nc}{N_{\!c}} 
\newcommand{\Nf}{N_{\!f}} 
\newcommand{\NA}{N_{\!A}} 
 
\newcommand{\MSbar}{\overline{\mbox{MS}}} 
\newcommand{\MOMbar}{\overline{\mbox{MOM}}} 

\begin{document}

\title{{\Large Exploring the infrared structure of QCD with the 
Gribov-Zwanziger Lagrangian}}

\author{J.A. Gracey}

\affiliation{Theoretical Physics Division, Department of Mathematical Sciences,
University of Liverpool, P.O. Box 147, Liverpool, L69 3BX, United Kingdom}


\begin{abstract}
We review recent one and two loop $\MSbar$ Landau gauge calculations using the 
Gribov-Zwanziger Lagrangian. The behaviour of the gluon and Faddeev-Popov ghost
propagators as well as the renormalization group invariant effective coupling 
constant is examined in the infrared limit.  


\end{abstract}

\maketitle

\thispagestyle{fancy}
\setcounter{page}{0}

\section{Introduction.}
Quantum Chromodynamics (QCD) is the quantum field theory describing the physics
of the strong nuclear force. In the ultraviolet r\'{e}gime the theory describes
asymptotically free quarks and gluons, \cite{1,2}, which at high energies 
behave as fundamental particles with propagators of the form $1/p^2$ as the 
momentum $p$ increases. However, free quarks and gluons are not observed in 
nature as isolated units due to the confinement mechanism. Moreover, there is a
wealth of evidence to indicate that at low energies the behaviour of the gluon
propagator differs significantly from that in the ultraviolet. (For example, 
see the recent reviews of \cite{3,4}.) For instance, both Dyson Schwinger 
equations (DSE) and lattice regularization observe that the gluon propagator 
vanishes at zero momentum, \cite{5,6,7,8,9,10}. This property is referred to as
gluon suppression. Whilst it has been established in the Landau gauge several 
other features emerge in the infrared. The propagator of the associated 
Faddeev-Popov ghost field diverges at zero momentum more rapidly than its 
$1/p^2$ ultraviolet behaviour. Again this ghost enhancement has been observed 
in DSE and lattice studies where both approaches obtain good agreement on the 
value of the exponent in the power law form of the propagator, \cite{5,6,7,8}. 
Another infrared property which is not fully resolved by DSE and lattice 
analyses is that of the freezing of a particular definition of the effective 
strong coupling constant. Whilst it is accepted that the particular quantity 
does not diverge at zero momentum, it is not clear whether it freezes to a 
finite or zero value. Given that these infrared properties of Landau gauge QCD 
have been well established by the numerically intensive DSE and lattice 
regularization techniques, one natural question to pose is, can such phenomena 
be probed from a Lagrangian point of view, and if not initially quantitatively,
can it be accessed and understood qualitatively? This is therefore the purpose 
of this article, which is primarily to review recent activity in this area, 
\cite{11,12}, by using Zwanziger's elegant study on the Gribov problem, 
\cite{13,14,15,16}.

Briefly, Gribov pointed out in \cite{17} that one cannot uniquely fix the gauge
globally in a non-abelian gauge theory. Specifically for some gauge
configurations one can construct one or more copies which can be related by a 
gauge transformation. To resolve this Gribov proposed that the path integral be 
restricted to the region defined by the first zero of the Faddeev-Popov 
operator which is defined as the Gribov volume. Whilst one can still have gauge
copies inside the Gribov horizon, \cite{18,19,20}, within it there is a smaller
region known as the fundamental modular region where there are no Gribov 
copies. The key feature of the restriction to the first Gribov region is the 
introduction of the Gribov mass scale, $\gamma$, which can be related to the 
volume of the first Gribov region. The restriction of the path integral 
significantly modifies the behaviour of the gluon and ghost propagators in the 
infrared, \cite{17}. For instance, the (non-local) Gribov Lagrangian contains a
gluon which is infrared suppressed and an enhanced Faddeev-Popov ghost. The 
latter derives from the one loop ghost $2$-point function where the gap 
equation satisfied by the Gribov mass is central to the {\em one} loop 
enhancement. Whilst Gribov's Landau gauge analysis gave significant insight 
into the infrared properties, one difficulty was the inability to easily 
perform practical computations since Gribov's approach effectively amounted to 
using a Lagrangian with a non-locality. The breakthrough in this area came with
Zwanziger's construction of a localised Lagrangian in \cite{13,14,15,16}. 
Significantly that Lagrangian is renormalizable, \cite{21,22}, and therefore 
allows one to carry out calculations. 

\section{Gribov-Zwanziger Lagrangian.} 
The Gribov-Zwanziger Lagrangian involves two pieces, \cite{14}. One is the 
usual Landau gauge fixed QCD Lagrangian and the other part involves additional 
fields which we will refer to as Zwanziger ghosts. These are $\{ \phi_\mu^{ab},
\bar{\phi}_\mu^{ab}, \omega_\mu^{ab}, \bar{\omega}_\mu^{ab} \}$ where the 
latter pair are anticommuting. The full localised renormalizable Lagrangian is,
\cite{14}, 
\begin{eqnarray}
L^{GZ} &=& L^{QCD} ~+~ \bar{\phi}^{ab \, \mu} \partial^\nu
\left( D_\nu \phi_\mu \right)^{ab} ~-~ \bar{\omega}^{ab \, \mu} \partial^\nu 
\left( D_\nu \omega_\mu \right)^{ab} \nonumber \\  
&& -~ g f^{abc} \partial^\nu \bar{\omega}^{ae}_\mu \left( D_\nu c \right)^b
\phi^{ec \, \mu} \nonumber \\
&& +~ \frac{\gamma^2}{\sqrt{2}} \left( f^{abc} A^{a \, \mu} \phi^{bc}_\mu ~+~ 
f^{abc} A^{a \, \mu} \bar{\phi}^{bc}_\mu \right) \,-\, 
\frac{d \NA \gamma^4}{2g^2} 
\label{laggz}
\end{eqnarray} 
where $L^{QCD}$ denotes the usual Lagrangian involving the gluon $A^a_\mu$ and
Faddeev-Popov ghost fields $c^a$ and $\bar{c}^a$, $D_\mu$ is the usual 
covariant derivative, $f^{abc}$ are the colour group structure functions, $\NA$
is the dimension of the adjoint representation, $g$ is the coupling constant, 
and $d$ is the spacetime dimension. The commuting Zwanziger ghosts 
$\phi_\mu^{ab}$ and $\bar{\phi}_\mu^{ab}$ implement the Gribov horizon 
condition which in the Gribov formulation is defined by the vacuum expectation 
value, \cite{17},  
\begin{equation} 
\left\langle A^a_\mu(x) \frac{1}{\partial^\nu D_\nu} A^{a\,\mu}(x) 
\right\rangle ~=~ \frac{d N_A}{C_A g^2} ~. 
\end{equation} 
The equivalent definition in (\ref{laggz}) is, \cite{14},  
\begin{equation} 
f^{abc} \langle A^{a \, \mu}(x) \phi^{bc}_\mu(x) \rangle ~=~ 
\frac{d \NA \gamma^2}{\sqrt{2}g^2} 
\label{gapdefgz} 
\end{equation} 
and these two equations define the gap equation satisfied by the Gribov mass 
$\gamma$. The relation between both definitions of the horizon conditions is
made manifest from the equation of motion of the Zwanziger ghost
\begin{equation} 
\phi^{ab}_\mu ~=~ \frac{\gamma^2}{\sqrt{2}} f^{abc} 
\frac{1}{\partial^\nu D_\nu} A^c_\mu ~. 
\label{eqnm} 
\end{equation} 
Further, focusing on the part of the Lagrangian involving purely gluons and 
using (\ref{eqnm}) then the non-locality of the original formulation of Gribov 
becomes apparent  
\begin{equation}
L ~=~ -~ \frac{1}{4} G_{\mu\nu}^a 
G^{a \, \mu\nu} ~+~ \frac{C_A\gamma^4}{2} A^a_\mu \, \frac{1}{\partial^\nu 
D_\nu} A^{a \, \mu} ~-~ \frac{d \NA \gamma^4}{2g^2} ~.  
\end{equation}

As with any Lagrangian one needs to determine the explicit values of the
renormalization constants and hence anomalous dimensions in order to perform 
practical calculations. As was demonstrated in \cite{21,22} the Landau gauge 
Gribov-Zwanziger Lagrangian has an interesting renormalization structure. In 
addition to the renormalization constants of the usual QCD Lagrangian the new 
fields and parameters each have a renormalization constant  
\begin{equation} 
\phi^{ab \, \mu}_{\mbox{o}} ~=~ \sqrt{Z_\phi} \, 
\phi^{ab \, \mu} ~~,~~ 
\omega^{ab \, \mu}_{\mbox{o}} ~=~ \sqrt{Z_\omega} \, 
\omega^{ab \, \mu} ~~,~~ 
\gamma_{\mbox{o}} ~=~ Z_\gamma \gamma
\end{equation} 
where the subscript ${}_{\mbox{o}}$ denotes the bare quantity. However, it was
demonstrated in \cite{21,22} by the algebraic renormalization programme that to 
{\em all} orders in perturbation theory in the Landau gauge
\begin{equation}
Z_c ~=~ Z_\phi ~=~ Z_\omega
\label{stid1}
\end{equation}
and 
\begin{equation}
Z_\gamma ~=~ Z_A^{1/4} Z_c^{1/4}
\label{stid2}
\end{equation}
where $Z_A$ and $Z_c$ are respectively the gluon and Faddeev-Popov ghost
renormalization constants. Hence, no new renormalization constants are required
and, moreover, the original QCD anomalous dimensions for the gluon, 
Faddeev-Popov ghost and quark as well as the $\beta$-function remain unchanged. 
So the presence of the localising fields does not alter the ultraviolet 
running. For completeness we note that the anomalous dimension of $\gamma$ is 
known to three loops in $\MSbar$ for arbitrary colour group and at four loops 
for $SU(\Nc)$ in the Landau gauge, \cite{23}. 

Whilst this article primarily deals with the Landau gauge, as an aside one can 
begin to address what happens in other linear covariant gauges aside from the 
Landau gauge by restoring the usual $\alpha$ dependent part to $L^{QCD}$ in
(\ref{laggz}). In \cite{24} the construction of the horizon function has been 
examined for small $\alpha$ where $\alpha$ is the gauge fixing parameter in a
linear covariant gauge. In this approximation one can repeat some of the Landau
gauge construction and establish a gap equation. In this context one can 
examine the renormalization of (\ref{laggz}) where $L^{QCD}$ is replaced by the
$\alpha$ dependent Lagrangian. It is a straightforward exercise to examine the 
three loop renormalization for non-zero $\alpha$ which has not been determined
before. The machinery to do this involves the application of the {\sc Mincer} 
algorithm, \cite{25,26}, written in the symbolic manipulation language {\sc 
Form}, \cite{27}, and the intensive use of computer algebra. The full three 
loop renormalization of the $\alpha$ extended Gribov-Zwanziger Lagrangian has 
been performed in the $\MSbar$ scheme and the following $\alpha$ dependent 
anomalous dimensions emerged for the additional fields and $\gamma$ to three
loops 
\begin{eqnarray} 
\gamma_c(a) &=& \gamma_\phi(a) ~=~ \gamma_\omega(a) \nonumber \\  
\gamma_\gamma(a) &=& [ 16 T_F \Nf - ( 35 + 3 \alpha ) C_A ] \frac{a}{48} 
\nonumber \\
&& +~ [ ( 3\alpha - 449 ) C_A^2 + 280 T_F \Nf C_A + 192 T_F \Nf C_A ] 
\frac{a}{192} \nonumber \\ 
&& +~ [ ( 1512 \alpha - 15552 \zeta(3) + 89008 ) T_F \Nf C_A^2 \nonumber \\ 
&& -~ 3456 T_F \Nf C_F^2 - 8448 T_F^2 \Nf^2 C_F - 12352 T_F^2 \Nf^2 C_A
\nonumber \\
&& +~ ( 20736 \zeta(3) + 19920 ) T_F \Nf C_A C_F 
\nonumber \\
&& -~ ( 81 \alpha^3 - 162 \zeta(3) \alpha^2 + 162 \alpha^2 - 648 \zeta(3) 
\alpha + 918 \alpha \nonumber \\
&& -~ 486 \zeta(3) + 75607 ) C_A^3 ] \frac{a^3}{6912} ~+~ O(a^4)  
\end{eqnarray} 
where $a$~$=$~$g^2/(16\pi^2)$, $\zeta(n)$ is the Riemann zeta function and the
group Casimirs are defined by
\begin{equation}
T^a T^a ~=~ C_F I ~~,~~ f^{acd} f^{bcd} ~=~ C_A \delta^{ab} ~~,~~
\mbox{Tr} ( T^a T^b ) ~=~ T_F \delta^{ab} 
\end{equation}
where $T^a$ are the generators of the Lie algebra of the colour group.
Throughout we use dimensional regularization in $d$~$=$~$4$~$-$~$2\epsilon$
dimensions. To establish $\gamma_\gamma(a)$ we have used the version of the
Gribov-Zwanziger Lagrangian where the $\gamma$ dependent terms are regarded as
being part of the interaction Lagrangian. Therefore, the gluon and 
$\phi^{ab}_\mu$ fields are regarded as massless. This is the form required for
applying the {\sc Mincer} algorithm which determines massless $2$-point 
functions to three loops and the finite part in $\epsilon$. Thus 
$\gamma_\gamma(a)$ is determined by renormalizing the Green's function where
the operator associated with $\gamma^2$ in (\ref{laggz}) is inserted at zero
momentum into the $A^a_\mu$-$\phi^{bc}_\nu$ Green's function. This involved $2$
one loop, $43$ two loop and $1082$ three loop Feynman diagrams which were
generated with the {\sc Qgraf} package, \cite{28}. 

Clearly the first Slavnov-Taylor identity, (\ref{stid1}), appears to hold for 
all $\alpha$ and is central to the observation that like the Landau gauge the 
other anomalous dimensions (and the $\beta$-function) are equivalent to their 
$\alpha$ dependent values with or without the localising features. By contrast 
the Gribov mass parameter is not only $\alpha$ dependent but is {\em not} 
proportional to the sum of the gluon and Faddeev-Popov ghost anomalous 
dimensions as was the case in the Landau gauge. Therefore, for non-zero 
$\alpha$ an extra independent renormalization constant is required to render 
(\ref{laggz}) finite unlike the Landau gauge. 
 
\section{Gap equation.} 
Having established the Gribov-Zwanziger Lagrangian as a tool to perform
calculations one can begin to address the problem of what its properties are at
zero momentum, returning to the Landau gauge. First, one can construct the gap 
equation satisfied by $\gamma$ at two loops by evaluating the vacuum 
expectation value (\ref{gapdefgz}). This requires the Feynman rules of 
(\ref{laggz}) as well as the propagators. Due to the mixed $2$-point term in 
(\ref{laggz}) the $A^a_\mu$ and $\phi^{ab}_\mu$ sector propagators are  
\begin{eqnarray}
\langle A^a_\mu(p) A^b_\nu(-p) \rangle &=& -~ 
\frac{\delta^{ab}p^2}{[(p^2)^2+C_A\gamma^4]} P_{\mu\nu}(p) \nonumber \\  
\langle A^a_\mu(p) \bar{\phi}^{bc}_\nu(-p) \rangle &=& -~ 
\frac{f^{abc}\gamma^2}{\sqrt{2}[(p^2)^2+C_A\gamma^4]} P_{\mu\nu}(p) 
\nonumber \\  
\langle \phi^{ab}_\mu(p) \bar{\phi}^{cd}_\nu(-p) \rangle &=& -~ 
\frac{\delta^{ac}\delta^{bd}}{p^2}\eta_{\mu\nu} \nonumber \\
&& +~ \frac{f^{abe}f^{cde}\gamma^4}{p^2[(p^2)^2+C_A\gamma^4]} P_{\mu\nu}(p) 
\end{eqnarray} 
where $P_{\mu\nu}(p)$ $=$ $\eta_{\mu\nu}$ $-$ $p_\mu p_\nu/p^2$. Therefore
using these to evaluate the $1$ one loop and $17$ two loop Feynman graphs
contributing to (\ref{gapdefgz}), we arrive at the two loop $\MSbar$ gap
equation satisfied by the running Gribov mass $\gamma$, for $\Nf$ massless 
quarks, \cite{11},  
\begin{eqnarray} 
1 &=& C_A \left[ \frac{5}{8} - \frac{3}{8} \ln \left( 
\frac{C_A\gamma^4}{\mu^4} \right) \right] a \nonumber \\ 
&+& \left[ C_A^2 \left( \frac{2017}{768} - \frac{11097}{2048} s_2
+ \frac{95}{256} \zeta(2)
- \frac{65}{48} \ln \left( \frac{C_A\gamma^4}{\mu^4} \right) \right. \right. 
\nonumber \\ 
&& \left. \left. +~ \frac{35}{128} \left( \ln \left( 
\frac{C_A\gamma^4}{\mu^4} \right) \right)^2 + \frac{1137}{2560} \sqrt{5} 
\zeta(2) - \frac{205\pi^2}{512} \right) \right. \nonumber \\
&& \left. +~ C_A T_F \Nf \left( -~ \frac{25}{24} - \zeta(2)
+ \frac{7}{12} \ln \left( \frac{C_A\gamma^4}{\mu^4} \right) \right. \right.
\nonumber \\ 
&& \left. \left. -~ \frac{1}{8} \left( \ln \left( 
\frac{C_A\gamma^4}{\mu^4} \right) \right)^2 + \frac{\pi^2}{8} \right) \right] 
a^2 +~ O(a^3) 
\label{gapeqn}
\end{eqnarray} 
where $s_2$ $=$ $(2\sqrt{3}/9) \mbox{Cl}_2(2\pi/3)$ with $\mbox{Cl}_2(x)$ the
Clausen function and $\mu$ is the scale introduced in dimensional 
regularization which ensures a dimensionless coupling constant in 
$d$-dimensions. This calculation also required the use of the {\sc Qgraf} 
package, \cite{28}, to generate the electronic version of the Feynman diagrams 
as well as the use of {\sc Form} to handle the algebra. Additionally the two 
loop master vacuum bubble integral  
\begin{equation}
I(m_1^2,m_2^2,m_3^2) ~=~ 
\int_{kl} \frac{1}{[k^2-m_1^2][l^2-m_2^2][(k-l)^2-m_3^2]}  
\label{masint} 
\end{equation}
where $m_i^2$~$\in$~$\{0, i \sqrt{C_A} \gamma^2, - i \sqrt{C_A} \gamma^2 \}$ 
was required to the finite part in $\epsilon$.

\section{Infrared properties.}
As Gribov was able to establish the one loop Faddeev-Popov ghost enhancement by
evaluating the ghost $2$-point function and applying the one loop gap equation,
\cite{17}, it is now viable to extend that calculation. At two loops there are 
$31$ Feynman diagrams to evaluate. If we write the Faddeev-Popov form factor as
\begin{equation}
D_c(p^2) ~=~ \frac{1}{[ 1 + u(p^2) ]}
\end{equation}
where $D_c(p^2)/p^2$ represents the propagator itself, then $u(p^2)$ represents
the radiative corrections. So ghost enhancement then translates into the 
Kugo-Ojima confinement criterion of $u(0)$~$=$~$-$~$1$, \cite{29}. Hence, to 
consider this in the Gribov-Zwanziger context one needs to evaluate the two 
loop radiative corrections in the small $p^2$ expansion. In practical terms 
this means carrying out the vacuum bubble expansion of the Feynman diagrams and
using the master integral (\ref{masint}), for instance, and the cases where the
propagators have integer powers larger than unity. Completing this one finds
that $u(0)$~$=$~$-$~$1$ at {\em two} loops {\em precisely} when $\gamma$
satisfies the gap equation of (\ref{gapeqn}), \cite{11}. In extending the one 
loop result of Gribov to two loops means that the Kugo-Ojima confinement 
criterion is consistent at two loops in (\ref{laggz}) as well as being a check 
on the actual explicit computation of (\ref{gapeqn}). Moreover, the fact that 
the condition is precisely satisfied at two loops is also consistent with the 
{\em all orders} proof given in \cite{15}. There general arguments established 
that the ghost propagator behaves as $1/(p^2)^2$ as $p^2$~$\rightarrow$~$0$ in 
(\ref{laggz}). 

Repeating the analysis for the gluon propagator is more involved due to the
mixing of $A^a_\mu$ with $\phi^{ab}_\mu$ in the quadratic part of the 
Lagrangian. If one formally writes the one loop correction to the 
$2$~$\times$~$2$ matrix of $2$-point functions as, \cite{12}, 
\begin{equation}
\left(
\begin{array}{cc}
p^2 \delta^{ac} & - \gamma^2 f^{acd} \\
- \gamma^2 f^{cab} & - p^2 \delta^{ac} \delta^{bd} \\
\end{array}
\right) \, + \, 
\left(
\begin{array}{cc}
X \delta^{ac} & U f^{acd} \\ M f^{cab} & N^{abcd} \\
\end{array}
\right) a ~+~ O(a^2) 
\label{mat2pt}
\end{equation}
where the common factor of $P_{\mu\nu}(p)$ in the Landau gauge has been 
factored off, $N^{abcd}$ $=$ $Q \delta^{ac} \delta^{bd}$ $+$ $W f^{ace} 
f^{bde}$ $+$ $R f^{abe} f^{cde}$ $+$ $S d_A^{abcd}$ and
\begin{equation} 
d_A^{abcd} ~=~ \mbox{Tr} \left( T^{(a}_A T^b_A T^c_A T^{d)}_A \right)
\end{equation} 
where $T^a_A$ is the colour group generator in the adjoint representation. The
$14$ one loop Feynman graphs contributing to the $2$-point function matrix have
been evaluated both exactly and in the vacuum bubble expansion and are in
agreement when the exact expression is expanded in powers of $p^2$. Inverting
the $2$-point matrix gives the propagator matrix and formally we find the
propagators to one loop are given by, \cite{12},  
\begin{eqnarray}
&& \left(
\begin{array}{cc}
\frac{p^2}{[(p^2)^2+C_A\gamma^4]} \delta^{cp} & 
- \frac{\gamma^2}{[(p^2)^2+C_A\gamma^4]} f^{cpq} \\
- \frac{\gamma^2}{[(p^2)^2+C_A\gamma^4]} f^{pcd} & 
- \frac{1}{p^2} \delta^{cp} \delta^{dq} 
+ \frac{\gamma^4 f^{cdr} f^{pqr}}{p^2[(p^2)^2+C_A\gamma^4]} \\
\end{array}
\right) \nonumber \\
&+& \left(
\begin{array}{cc}
A \delta^{cp} & C f^{cpq} \\
E f^{pcd} & \Phi^{cdpq} \\
\end{array}
\right) a ~+~ O(a^2) 
\end{eqnarray} 
where
\begin{equation}
\Phi^{cdpq} ~=~ G \delta^{cp} \delta^{dq} + J f^{cpe} f^{dqe} 
+ K f^{cde} f^{pqe} + L d_A^{cdpq} 
\end{equation} 
and 
\begin{eqnarray}
A &=& \frac{(-1)}{[(p^2)^2+C_A\gamma^4]^2} \left[ (p^2)^2 X 
- 2 C_A \gamma^2 p^2 U \right. \nonumber \\
&& \left. ~~~~~~~~~~~+~ C_A \gamma^4 \left( Q + C_A R + \half C_A W \right)
\right] ~. 
\end{eqnarray}
The first entry in this matrix corresponds to the correction to the gluon
propagator and in terms of the $2$-point functions is 
\begin{eqnarray}
A &=& \frac{(-1)}{[(p^2)^2+C_A\gamma^4]^2} \left[ (p^2)^2 X 
- 2 C_A \gamma^2 p^2 U \right. \nonumber \\
&& \left. ~~~~~~~~~~~+~ C_A \gamma^4 \left( Q + C_A R + \half C_A W \right)
\right] ~.  
\end{eqnarray}
Hence, if one writes the form factor of the gluon as $D_A(p^2)$ where $D_A(p^2)
P_{\mu\nu}(p)/p^2$ is the propagator, then one finds that 
$D_A(p^2)$~$\rightarrow$~$0$ as $p^2$~$\rightarrow$~$0$ at one loop, \cite{12}.
Hence, one has gluon suppression at this order in (\ref{laggz}) consistent with
DSE and lattice studies.

We close this section on the propagators by noting that for the other fields
one and two loop calculations reveal parallel behaviour. For the 
$\omega^{ab}_\mu$ field it too has a completely similar enhancement at two 
loops if one computes the two loop $2$-point function, in a similar way to that
of the Faddeev-Popov ghost, and applies the two loop gap equation. So the 
Kugo-Ojima confinement criterion is satisfied by {\em both} sets of 
anticommuting fields. However, it is worth noting that the original Kugo-Ojima 
analysis was for a non-abelian gauge theory with only Faddeev-Popov ghosts and 
not with the additional Zwanziger ghosts. Therefore, for completeness it would 
seem that the study of \cite{29} needs to be extended to the case of 
(\ref{laggz}). Equally at one loop the {\em exact} expressions for both the 
$\omega^{ab}_\mu$ and Faddeev-Popov ghost form factors are the same. Explicitly
we have 
\begin{eqnarray}  
&& D_c(p^2) ~=~ D_\omega(p^2) \nonumber \\
&&= \left[ \left[ \frac{5}{8} ~+~ 
\frac{\pi p^2}{8\sqrt{C_A}\gamma^2} ~+~ \frac{3 \sqrt{C_A} \gamma^2}{4p^2} 
\tan^{-1} \left[ \frac{\sqrt{C_A}\gamma^2}{p^2} \right] 
\right. \right. \nonumber \\
&& \left. \left. ~~-~ \frac{3}{8} \ln \left[ 1 + \frac{(p^2)^2}{C_A\gamma^4} 
\right] ~+~ \frac{C_A\gamma^4}{8(p^2)^2} \ln 
\left[ 1 + \frac{(p^2)^2}{C_A\gamma^4} \right] \right. \right. \nonumber \\
&& \left. \left. ~~-~ \frac{3 \pi \sqrt{C_A} \gamma^2}{8p^2} 
~-~ \frac{p^2}{4\sqrt{C_A}\gamma^2} \tan^{-1} \left[ 
\frac{\sqrt{C_A}\gamma^2}{p^2} \right] \right] C_A a \right]^{-1} ~.
\nonumber \\  
\end{eqnarray}  
Moreover, these facts suggest that there is an all orders equality which, if 
so, could possibly be established by general symmetry arguments. Also, whilst 
the $\phi^{ab}_\mu$ $2$-point function is more complicated due to the 
appearance of the colour group tensors, the channel corresponding to the tree 
term in the Lagrangian is equivalent at one loop in $\MSbar$ to the gap 
equation. This suggests that in this channel there is a change in its infrared 
behaviour.
 
\section{Effective coupling constant.}
One quantity which receives wide attention in both DSE and lattice studies is
the renormalization group invariant definition of the effective strong coupling
constant based on the renormalization properties of the ghost gluon vertex.
(See, for instance, \cite{4}.) Since this vertex does not undergo 
renormalization in the Landau gauge, then the effective coupling constant is 
defined from the gluon and ghost form factors as 
\begin{equation}
\alpha^{\mbox{eff}}_S (p^2) ~=~ \alpha(\mu) D_A(p^2) \left(
D_c(p^2) \right)^2 
\label{aldef}
\end{equation}
where $\alpha(\mu)$~$=$~$g^2(\mu)/(4\pi)$ is the running strong coupling
constant. Although both numerical approaches use the same definition they do 
not definitively agree on the infrared behaviour aside from the fact that there
is freezing at a {\em finite} value. However, some DSE studies find 
$\alpha^{\mbox{eff}}_S (0)$~$=$~$2.97$ for $SU(3)$, \cite{30,31}, whilst 
various lattice and DSE analyses suggest $\alpha^{\mbox{eff}}_S (0)$~$=$~$0$, 
\cite{32,33}. Given that we have analysed both form factors in the 
Gribov-Zwanziger Lagrangian, it is a simple exercise to evaluate (\ref{aldef}) 
in the $p^2$~$\rightarrow$~$0$ limit. We find that at one loop in the $\MSbar$ 
scheme, \cite{12}, 
\begin{equation}
\alpha^{\mbox{eff}}_S (0) ~=~ \lim_{p^2 \rightarrow 0} \left[ 
\frac{ \alpha(\mu) \left[ 1 + C_A \left( \frac{3}{8} L - \frac{215}{384} 
\right) a \right] (p^2)^2 }
{ C_A \gamma^4 \left[ 1 + C_A \left( \frac{3}{8} L - \frac{5}{8} 
+ \frac{\pi p^2}{8 \sqrt{C_A} \gamma^2} \right) a \right]^2 } \right] 
\end{equation}
where $L$ $=$ $\ln \left( \frac{C_A \gamma^4}{\mu^4} \right)$. Thus for $SU(3)$ 
we find $\alpha^{\mbox{eff}}_S (0)$~$=$~$1.768$ independent of the number of 
massless quarks. Whilst this is less than the value for the DSE approach it is 
finite and non-zero. Moreover, the $\Nc$ dependence is in qualitative agreement
with \cite{30,31}. In the determination of a finite value it is worth noting 
the key role played by the gap equation in ensuring the cancellation of the 
dimensionful and dimensionless quantities to leave a pure number. We also 
remark that the $\omega^{ab}_\mu$ gluon vertex has a similar freezing to the 
{\em same} value, \cite{12}. 

One property of (\ref{aldef}) which motivated recent studies of Landau gauge
QCD and the suggested condensation of low dimensional operators was the 
numerical evidence of a $1/p^2$ power correction to the effective coupling 
constant, \cite{34,35,36,37}. Although that analysis was in the $\MOMbar$ 
scheme, if it is a more general property of the Landau gauge effective coupling
constant it ought to be present in other schemes. As we have exact expressions 
for the form factors it is possible to evaluate (\ref{aldef}) in powers of 
$p^2$ and see whether a $1/p^2$ term emerges as a correction or if the expected
$1/(p^2)^2$ term is the first term. The latter is certainly expected to be 
present due to the existence of the gauge invariant dimension four operator 
$(G^a_{\mu\nu})^2$ which can condense. Therefore, returning to (\ref{mat2pt}) 
and formally expanding in powers of $\gamma^2$, we have at leading order at one
loop, \cite{12},  
\begin{eqnarray}
X &=& \left[ \left[ \left( \frac{13}{6} \ln \left( \frac{p^2}{\mu^2} \right) 
- \frac{97}{36} \right) p^2 + \frac{3\pi\sqrt{C_A}\gamma^2}{8} \right] C_A 
\right. \nonumber \\
&& \left. ~-~ \left[ \frac{4}{3} \ln \left( \frac{p^2}{\mu^2} \right) 
- \frac{20}{9} \right] T_F \Nf p^2 \right] a \nonumber \\  
M &=& U ~=~ \left[ \frac{11C_A}{8} \gamma^2 \right] a \nonumber \\ 
Q &=& \left[ \left[ ~-~ \left( 1 - \frac{3}{4} \ln \left( \frac{p^2}{\mu^2} 
\right) \right) p^2 ~+~ \frac{3\pi\sqrt{C_A}\gamma^2}{8} \right] C_A \right] a 
\nonumber \\
W &=& R ~=~ S ~=~ O(a^2) ~. 
\end{eqnarray} 
Hence, at one loop  
\begin{eqnarray}
D_A(p^2) &=& 1 ~-~ \frac{3C_A\pi}{8} \frac{\sqrt{C_A}\gamma^2}{p^2} a
\\ 
D_c(p^2) &=& \left[ -\, 1 + \left[ 1 - \frac{3}{4} \ln \left( \frac{p^2}{\mu^2}
\right) - \frac{3\pi}{8} \frac{\sqrt{C_A}\gamma^2}{p^2} \right] C_A a 
\right]^{-1} ~. \nonumber  
\end{eqnarray} 
Thus evaluating $\alpha^{\mbox{eff}}_S (p^2)$ in the same expansion and 
identifying the $\gamma$ independent pieces as deriving from the perturbative
part of the coupling constant, we find at one loop  
\begin{equation}
\alpha^{\mbox{eff}}_S (p^2) ~=~ 
\alpha^{\mbox{pert}}_S (p^2) \left[ 1 ~-~ \frac{9C_A\pi}{8}
\frac{\sqrt{C_A}\gamma^2}{p^2} a \,+\, O \left( \frac{1}{(p^2)^2} \right) 
\right] \,. 
\label{ccpow}
\end{equation}
Thus a $1/p^2$ correction emerges and the power correction of 
\cite{34,35,36,37} can be mimicked by the presence of the Gribov mass in the 
Gribov-Zwanziger Lagrangian. Though this ought to be qualified by noting that 
the condensation of the gauge variant operator $\half (A^a_\mu)^2$ has also 
received a large amount of attention in recent years in the Landau gauge, 
\cite{38,39}, and could be an alternative justification for the lattice result.
It is also worth observing that both $\gamma^2$ and $\half (A^a_\mu)^2$ have 
the {\em same} anomalous dimensions in the Landau gauge \cite{40,41}. Indeed 
the effect that the Gribov mass has on the condensation of $\half (A^a_\mu)^2$ 
has been examined in \cite{22,42,43}. One final point is that the sign of 
$\gamma^2$ is not fixed in (\ref{laggz}). Therefore, one can alter the sign of 
$\gamma^2$ in (\ref{ccpow}) to obtain the same sign as that in \cite{36,37}. 
Although this is for a different scheme ($\MOMbar$) compared to our $\MSbar$ 
result, it would result in an effective tachyonic gluon mass. This may appear 
unacceptable. However, as the gluon is not truly an observable field due to 
confinement this ought not to be a hindrance. More specifically this tachyonic 
nature is only revealed in the next to high energy limit. If one examines the 
full gluon propagator at one loop, there is no pole in the propagator. 
Therefore, one cannot truly identify a mass for the confined gluon in the 
conventional way that one does for a fundamental unconfined field. Another 
observation from a phenomenological point of view is that in \cite{44} it was 
noted that a tachyonic gluon mass in current correlators could reproduce 
experimental data more accurately than either a massless gluon or one with a 
non-tachyonic mass.

\vspace{0.4cm} 
\section{Conclusion.} 
We have studied the Gribov-Zwanziger Lagrangian, \cite{14}, at one loop and 
established that it qualitatively reproduces the propagator and effective 
coupling constant behaviour observed in the infrared using DSE and lattice 
studies. Moreover, the gap equation has been determined at two loops in 
$\MSbar$ and ensures the Kugo-Ojima confinement criterion is satisfied. This 
would appear to establish the Gribov-Zwanziger Lagrangian as a useful tool to 
examine other phenomenological aspects of QCD such as the operator product 
expansion and the evaluation of vacuum expectation values of gauge invariant 
operators, such as $(G^a_{\mu\nu})^2$. Further, we have examined the 
renormalization of the Gribov-Zwanziger Lagrangian in linear covariant gauges 
albeit in the limit of small gauge fixing parameter. Clearly it would be a 
formidable task to fully extend the Landau gauge Gribov-Zwanziger Lagrangian to
other gauges such as linear covariant and maximal abelian gauges, \cite{24,45}.
However, if it were possible then it would be interesting to extend our present
Landau gauge propagator and effective coupling constant analysis to these 
gauges in order to understand what their infrared structure is. This would 
complement DSE and lattice analyses.  

\section*{ACKNOWLEDGEMENTS.}
The author thanks Prof. S. Sorella, Prof. D. Zwanziger, and Dr D. Dudal for 
useful discussions concerning the Gribov problem.

\end{document}